\documentclass[%
 reprint,
 amsmath,amssymb,
 aps,floatfix,superscriptaddress,
nofootinbib]
{revtex4-2}

\usepackage{graphicx}
\usepackage{dcolumn}
\usepackage{soul} 
\usepackage{bm}
\usepackage{epstopdf}
 \usepackage[usenames,dvipsnames]{pstricks}
 \usepackage{epsfig}
 \usepackage{pst-grad} 
 \usepackage{pst-plot} 
 \usepackage[space]{grffile} 
 \usepackage{etoolbox} 
 \makeatletter 
 \patchcmd\Gread@eps{\@inputcheck#1 }{\@inputcheck"#1"\relax}{}{}
 \makeatother
\usepackage{booktabs}
\usepackage{boxhandler}
\usepackage{amsmath}
\usepackage{subcaption}
\usepackage{braket}
\usepackage[cmintegrals]{newtxmath}
\usepackage{multirow}
\usepackage{enumitem}

\begin{document}

\title{Performance enhancement of surface codes via recursive MWPM decoding}

\author{Antonio deMarti iOlius }
\email{ademartio@tecnun.es}
\affiliation{Department of Basic Sciences, Tecnun - University of Navarra, 20018 San Sebastian, Spain.}
\author{Josu Etxezarreta Martinez}
\affiliation{Department of Basic Sciences, Tecnun - University of Navarra, 20018 San Sebastian, Spain.}
\affiliation{These authors contributed equally.}
\author{Patricio Fuentes}
\affiliation{Department of Basic Sciences, Tecnun - University of Navarra, 20018 San Sebastian, Spain.}
\affiliation{These authors contributed equally.}
\author{Pedro M. Crespo}
\affiliation{Department of Basic Sciences, Tecnun - University of Navarra, 20018 San Sebastian, Spain.}

\date{\today}

\begin{abstract}
  
The minimum weight perfect matching (MWPM) decoder is the standard decoding strategy for quantum surface codes. However, it suffers a harsh decrease in performance when subjected to biased or non-identical quantum noise. In this work, we modify the conventional MWPM decoder so that it considers the biases, the non-uniformities and the relationship between $X$, $Y$ and $Z$ errors of the constituent qubits of a given surface code. Our modified approach, which we refer to as the recursive MWPM decoder, obtains an $18\%$ improvement in the probability threshold $p_{th}$ under depolarizing noise. We also obtain significant performance improvements when considering biased noise and independent non-identically distributed (i.ni.d.) error models derived from measurements performed on state-of-the-art quantum processors. In fact, when subjected to i.ni.d. noise, the recursive MWPM decoder yields a performance improvement of $105.5\%$ over the conventional MWPM strategy and, in some cases, it even surpasses the performance obtained over the well-known depolarizing channel.

\end{abstract}

\maketitle

\section{Introduction}

Quantum computers have the potential to surpass conventional machines and result in vast leaps forward in several fields such as cryptography, pharmacy or finance. Nonetheless, if quantum computers are to reach their, as of yet, unrealized potential, they must become fault-tolerant; they must be able to function in the presence of errors and imperfect components. The building blocks of quantum computers are known as qubits, and they are the most basic elements capable of storing quantum information. Thus, substantial effort and resources are currently being invested to design and build efficient methods to preserve information stored in qubits. This type of research falls within the scientific niche known as Quantum Error Correction (QEC). 

Recent developments in the field of QEC have revolved around the casting and adaptation of state-of-the art classical codes, such as the LDPC codes \cite{bicycle, qldpc15, patrick, reviewPat} and Turbo Codes \cite{josu1, josu2}, into the quantum framework with the goal of deriving powerful QEC codes. However, these schemes face significant technological limitations, such as the need for non-local interaction as large physical qubit counts, and so it is unlikely that they will be employed to implement near term quantum error correction. Fortunately, well-known strategies like the surface code \cite{surfacecodes1, surfacecodes2}, which have low connectivity and physical qubit counts, are promising candidates to protect quantum errors in the short term \cite{googlesurface, walliesurface}.


Information states encoded using surface codes are generally decoded using the Minimum Weight Perfect Matching decoder (MWPM), although a variety of different methods, like the Union Find decoder \cite{uf}, the Matrix Product State decoder \cite{mps}, the Renormalization Group decoder \cite{renormalization}, the Cellular Automata decoder \cite{automata}, the Neural Network decoder \cite{neural}, or even the recently-proposed `Belief Propagation (BP) + Ordered Statistics Decoder (OSD)' decoder \cite{bposd, osd2}, exist. When a quantum information state encoded within a surface code suffers an error, the MWPM method maps the resulting nonzero syndrome elements into a graph and then determines the set of edges without common vertices, referred as matching, in which the sum of weights is minimum \cite{mwpm, localmwpm}. The term perfect refers to the fact that the matching includes all vertices of the graph.

In this paper we introduce a modified MWPM approach which yields significant performance improvements when compared to the conventional method. We make use of the ideas proposed in \cite{xzcorrelation, fowlercorr}, where the authors exchange information between CSS code subgraphs, and in \cite{oldrec} where the idea of applying such concepts in a recursive matter under depolarizing noise is introduced. Based on these results, we develop a strategy that makes the rotated planar code more robust towards quantum noise. We verify this outcome by testing the performance of the rotated planar code over three different noise models: symmetric depolarizing noise, biased noise, and independent non-identically distributed (i.ni.d.) noise built using data from state of the art quantum processors \cite{googlesurface, Aspen, Zuchongzhi, Wash}.  


\section{The surface code}

Surface codes are a widely-studied class of quantum error correcting codes that are generally described by arranging data qubits in a 2-dimensional lattice. Data qubits interact locally with other additional qubits (check or ancilla qubits) that can later be measured to produce a syndrome. Several lattice structures have been studied for surface codes, but the most popular and the first one to be introduced is the square lattice \cite{surfacecodes1, surfacecodes2}.

Planar codes encode the information stored in a single logical qubit within larger sets of data qubits, which are located on the edges of the square lattice. On  the vertices and plaquettes of the code lattice additional qubits, known as measurement qubits, are found. Measurement qubits interact with their nearest data qubit neighbours differently depending on their location. This interaction occurs through a series of Hadamard and CNOT gates, allowing Pauli errors within the adjacent data qubits to propagate to the measurement qubit. Upon measurement, the measurement qubit brings partial information of their surrounding data qubits. This process can also be seen as measuring qubits applying Pauli operators to their adjacent data qubits and turning non-trivial if an odd number of adjacent data qubits undergo a Pauli error which anti-commutes with the operator applied by the measaurement qubit.For instance, so-called $X$-measurement qubits are located within the vertices of the planar code and apply $X$-operators to their neighbours, whereas $Z$-measurement qubits are located on the plaquettes and apply $Z$-operators in the same manner. Following their application on the corresponding data qubits, the measurement qubits themselves are measured. 
 Thus, the number of data qubits is restricted by the size of the planar code via the relation $n = d^2 + (d-1)^2$, where $n$ is the number of data qubits, and $d$ is the distance of the planar code. Moreover, the number of measurement qubits is $n-1$, since the planar code encodes a single qubit. The distance $d$ of the code represents the minimum number of Pauli gates that need to act on an encoded state in order to modify it non-trivially.

The planar code is initialized by establishing all constituent physical qubits into a desired state e.g. a tensor product of $\ket{0}$s and then measuring all the measurement qubits and saving their values. Note that the selection of the initial states of the physical qubits is not arbitrary in general since its selection impacts code performance whenever circuit-level noise is considered \cite{fragile}. However, since the present study does not consider noisy gates and SPAM (state preparation and measurement) errors, such selection is not important. Afterwards, whenever an error correction procedure is to be executed, the syndrome is extracted from the planar code by repeating the aforementioned procedure, i.e, measuring all the measurement qubits. Every measurement qubit that experiences a change in its measured outcome indicates that an odd number of nearest neighbours have experienced a Pauli error. $X$-measurement qubits are susceptible to $Z$-operators, $Z$-measurement qubits are susceptible to $X$-operators and both are susceptible to $Y$-operators, since in the effective Pauli group $Y=XZ$ \cite{reviewPat}. Therefore, we can say that the $X$-measurement qubits act as $Z$-checks while the $Z$-measurement qubits act as $X$-checks.


Against this backdrop, the planar code provides an effective way to encode the information of a qubit. It must be mentioned, however, that the current standard method employed to build experimental surface codes is the rotated planar code \cite{walliesurface, googlesurface}. The rotated planar code is a variant of the general planar code that reduces the number of physical qubits to $d^2$ by considering a rotated form of the original square lattice. This modified lattice results in a decrease of the code threshold \footnote{Assuming perfect measurements.} from around $16\%$ to $14\%$ through conventional decoding \cite{thresholdchange}, but it leads to an increase in the code rate, i.e, the ratio of logical qubits to physical qubits increases.

\begin{figure}[h]
    \centering
    \includegraphics[width=0.8\columnwidth,scale=1]{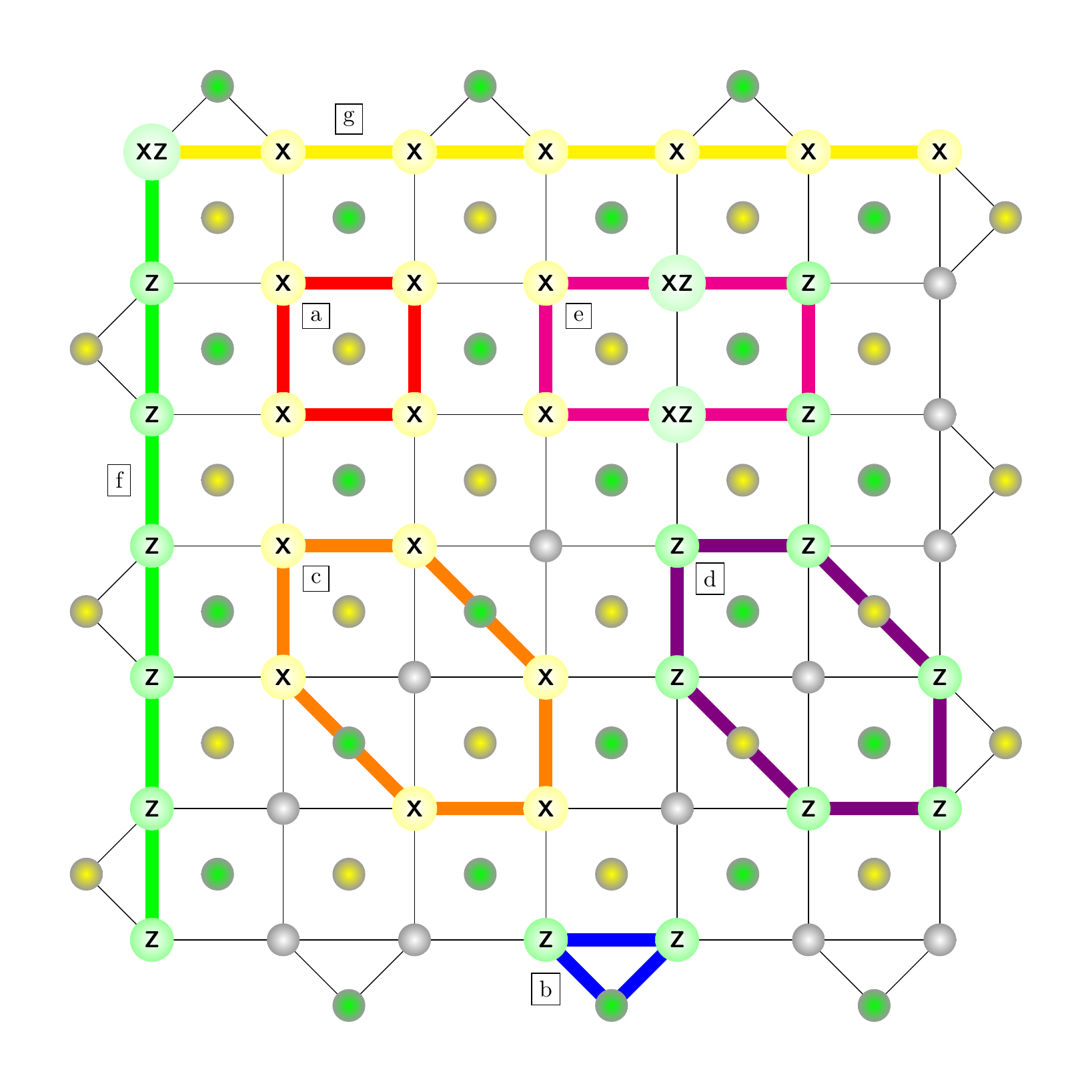}
    \caption{Visual representation of a $7\times7$ rotated planar code. The  data qubits, $X$-measurement, and $Z$-measurement qubits are represented via grey, yellow, and green circles respectively. In \textbf{a} we see an $X$-measurement qubit stabilizing its $4$ nearest data qubits. In \textbf{b} a $Z$-measurement qubit interacts with its two nearest data qubits. In \textbf{c} and \textbf{d} two adjacent $X$-measurement and $Z$-measurement qubits interact with their nearest data qubits. In \textbf{e}, an $X$-measurement stabilizes its nearest data qubits altogether with its adjacent $Z$-measurement qubit. The logical $X$ and $Z$-operators are shown in \textbf{g} and \textbf{f}  respectively. }
    \label{rotated planar}
\end{figure}

The rotated planar code is stabilized in the same way as the planar code: $X$-measurement and $Z$-measurement qubits are constantly initialized and measured. In Fig.\ref{rotated planar}, we portray the most relevant features of the planar code. As can be seen, the $X$ and $Z$-measurement stabilizers have weight $2$ or $4$, depending on their location within the code. Moreover, all adjacent $X$-measurement and $Z$-measurement commute amongst each another. Two measurement qubits of the same kind will commute because they apply the same type of operator to the shared data qubit. Moreover, as seen in label $e$, different kind measurements which are adjacent to each other also commute, as they anti-commute on two shared data qubits. Lastly, the logical operators $\hat{X}$ and $\hat{Z}$ can be understood as chains of $X$ and $Z$ operators traversing the code and commuting with all measurement qubits. The combination of both $\hat{X}$ and $\hat{Z}$ results in $\hat{Y}$ (notice that $\hat{X}\hat{Z} = -\hat{Z}\hat{X}$ through the anticommutation of $X$ and $Z$ in the top left corner data qubit).

An error within the code will result in a syndrome after a measurement of the stabilizer qubits. In order to extract a recovered error through said syndrome a process named decoding is undergone. Usually decoding requires the full syndrome, nevertheless some studies have been made as to optimize the syndrome data required by the decoder \cite{predecoder}. Minimum weight perfect matching is the conventional choice, nevertheless, this method suffers a harsh decrease in performance under realistic noise. 

\section{ The MWPM decoder}

The minimum weight perfect matching decoder allows us to decode quantum states encoded in planar codes by mapping non-zero syndrome elements to two separate subgraphs and matching them by finding the minimum weight configuration \cite{mwpm}. Each subgraph is composed by the $Z$-measurement and the $X$-measurement qubits.

\begin{figure}[h]
    \centering
    \includegraphics[width=0.6\columnwidth,scale=1]{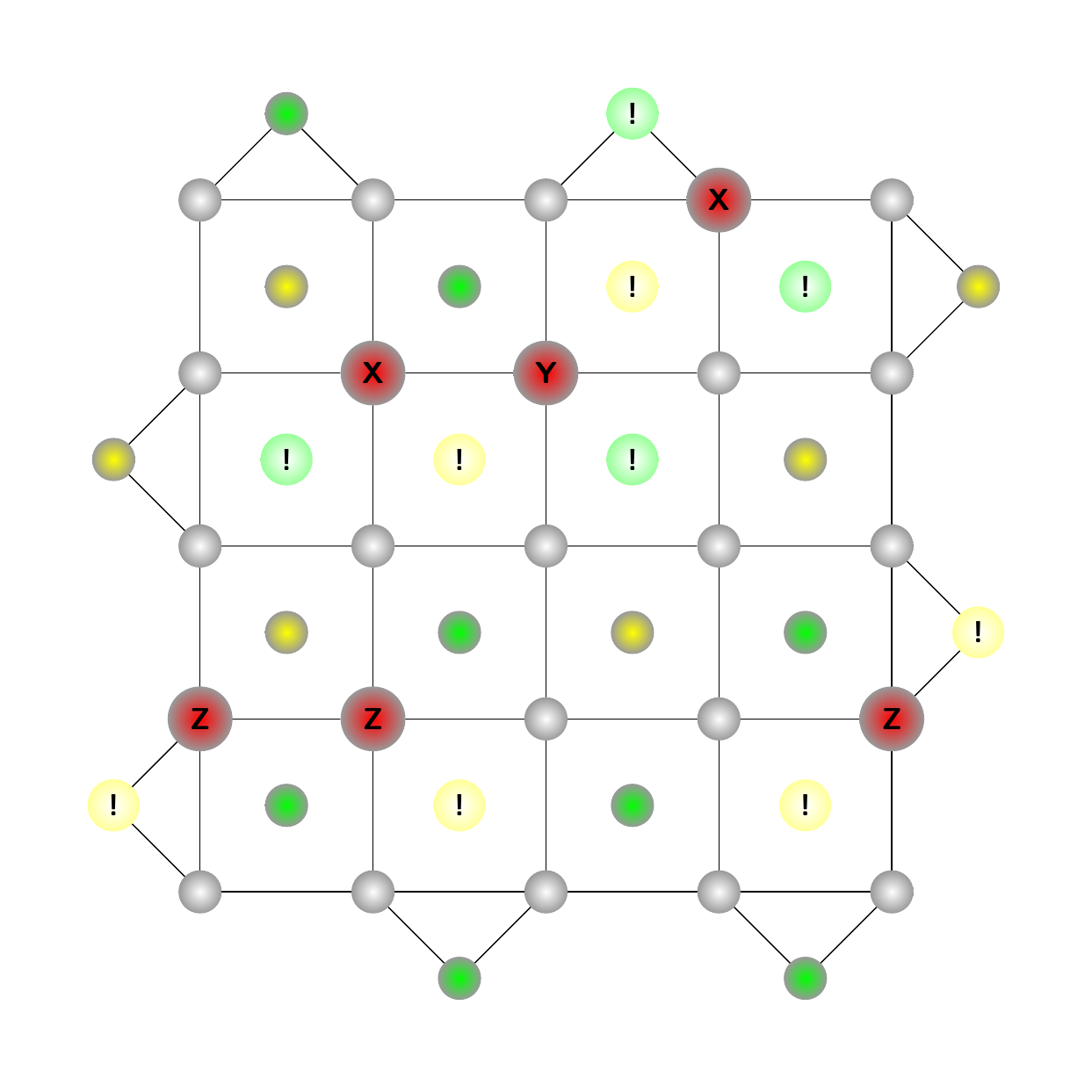}
    \vskip 0.1cm    
    \includegraphics[width=0.4\columnwidth,scale=1]{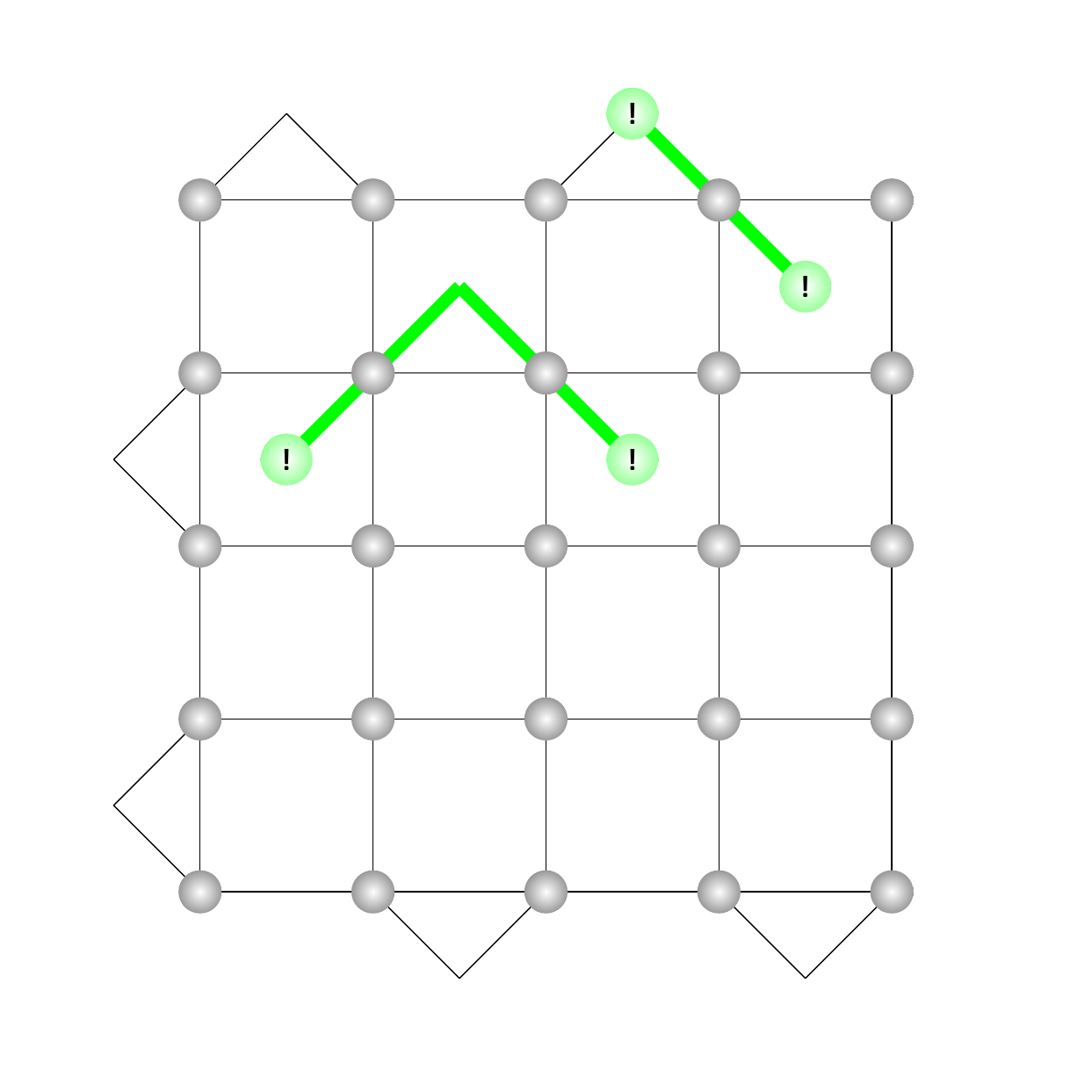}
    \includegraphics[width=0.4\columnwidth,scale=1]{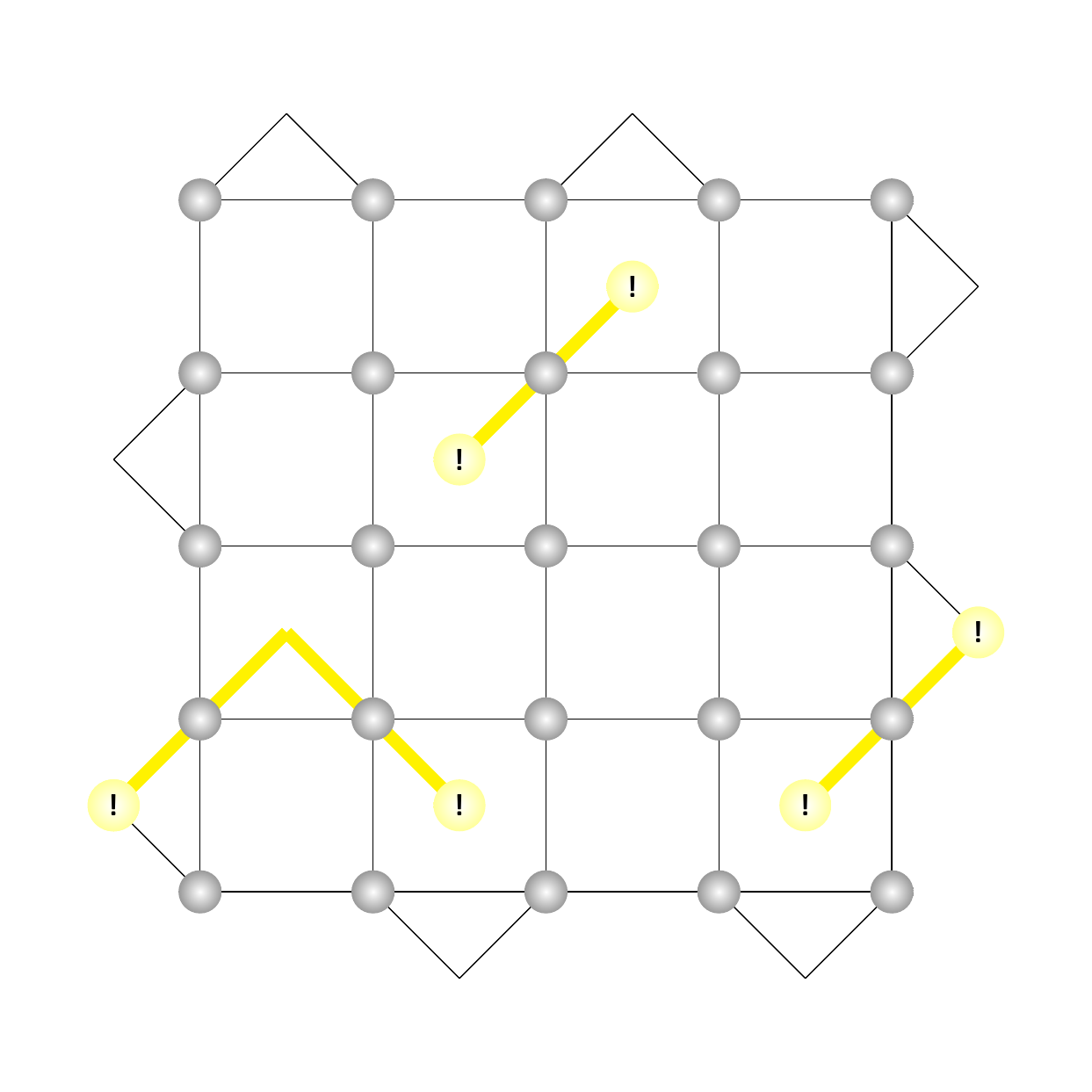}
    \caption{ Decoding of a $5\times5$ rotated planar code via minimum weight perfect matching. On the top, the rotated planar code experiences an error, the non-trivial Pauli operators on the data qubits are represented through red circles. Measurement qubits which experience a change in measurement due to the error are denoted in light green and light yellow circles with an exclamation mark. On the bottom, the $X$-check and $Z$-check subgraphs with the resulting MWPM computation are shown.}
    \label{MWPM}
\end{figure}

In Fig.\ref{MWPM}, an example of the functioning of the MWPM decoder is shown. On the top, the rotated planar code experiences an error, which is detected by a change in the measurement outcome of some measurement qubits. The measurement qubits are susceptible to errors that occur on their nearest data qubits: notice how the green $Z$-measurement qubits react to nearby $X$-errors, while the yellow $X$-measurement qubits react to $Z$-errors that are close. Additionally, both types of measurement qubits react to $Y$-errors, since they anticommute with both $X$ and $Z$-operators. Nevertheless, when a measurement qubit is in contact with an even number of errors with which it anticommutes, the anticommutations cancel out and it is never triggered. This effect can be seen in the yellow qubit in between the two $Z$-errors on the bottom left part of the graph, or with the green measurement qubit in between the $X$ and $Y$-errors. On the bottom of Fig.\ref{MWPM}, two subgraphs are created from the measurement qubits that experienced a change in their measurement outcome. Said measurement qubits will act as nodes on the resulting graph and the data qubits of the code are represented by the same weight edges which connect them. Once these two subgraphs are derived, the minimum weight perfect matching is computed, as is shown by the green and yellow lines of Fig.\ref{MWPM}. The graph paths that correspond to the MWPM represent the recovered error and so the recovery operation is built by applying $X$ and $Z$-operators on the data qubits that are crossed by green or yellow lines, respectively.

One of the strong suits of the MWPM decoding rule is that it always returns an estimate of the error that corresponds to the measured syndrome. If the MWPM decoder is successful but the input error is different from the recovered one, this difference will be up to a stabilizer element, and so the error correction procedure will still work. 


The characteristics of the MWPM decoding rule and its consideration of the degeneration of quantum states within the code makes it so that it can find the most likely recovery operation (one of the most likely errors belongs to the coset with highest probability \cite{reviewPat}).


while the average physical error probability stays below $10.3\%$ for each subgraph \cite{localmwpm}. For higher physical error probabilities, the most likely Pauli sequence no longer belongs to the stabilizer coset with highest probability, and so the performance will deteriorate until reaching the probability threshold ($p_{th}$): the physical error probability at which increasing the distance of the code no longer improves its performance.

Despite these positive traits, one of the biggest disadvantages of the MWPM decoder is its complexity, which scales with the distance of the rotated planar code as $O(d^6log(d))$ \cite{localmwpm}. It is likely that for large distance rotated planar codes other decoders will be applied, such as the local version of the MWPM or the so-called BPOSD decoder, as they significantly reduce complexity and generally achieve relatively similar performance results \cite{localmwpm,bposd, osd2}.

\section{The Recursive MWPM decoder}

As introduced in the previous sections, in this paper we consider the rotated planar code. Within it, the stabilizer checks of a conventional rotated planar code can be divided in two types: the $X$-checks, which interact by applying $Z$-gates to their nearest data qubit neighbours, and the $Z$-checks, which also interact with their nearest data qubit neighbours albeit via the application of $X$-gates. When errors take place, measuring the measurement qubits of the surface code produces a quantum syndrome, which is a binary vector that yields information regarding the error that has taken place. The process through which the most likely error to have caused the measured syndrome can be inferred is known as decoding. The most common decoding method applied with the surface code is the MWPM \cite{thresholdchange}, which, in broad terms consists in mapping the non-trivial syndrome elements to a graph, and then finding the minimum weight perfect matching associated to said graph (note that this yields the most probable error sequence, not the most likely error coset).

Conventional MWPM decoding performs well under depolarizing noise, but it suffers substantial performance losses when subjected to biased quantum noise \cite{tuckbias1, tuckbias2, xzzx}. It is noteworthy that this type of biased noise has similar effects on Belief Propagation (BP) decoders for QLDPC codes \cite{patrick2, roffe-bias}. The MWPM decoder suffers further decrements over more precise noise models like the earlier described i.ni.d. Pauli channel, where constituent qubits are considered individually and have their own probability distributions. In fact, when attempting to decode high standard deviation i.ni.d. noise with the conventional MWPM method harsh performance decreases are observed \cite{ton, fastfading}.

The conventional MWPM decoder assumes that the errors from each of the subgraphs are independent. However, they are correlated due to the fact that $Y$-errors are composed by a product of $X$ and $Z$-errors. The recursive MWPM (recMWPM) is an effort to take this extra information into account. More specifically, the rates $p_Y/p_X$ and $p_Y/p_Z$ are used to improve the performance under i.ni.d. noise. Let us see how. Consider the two check-subgraphs used for MWPM decoding. First, select whichever subgraph has the least non-zero syndrome elements. Were we to select the  $X$-check subgraph, once the MWPM is computed, the recovered error would indicate which data qubits are expected to have experienced $X$-errors. This recovered error will not only cover the set of $X$-errors experienced by the code, but also the set of $Y$-errors, since they also anti-commute with the $X$-checks. The $Y$-errors also anti-commute with the $Z$-checks, thus, each $X$-error recovered in the $X$-check subgraph should also appear in the $Z$-check one with probability: $\frac{p_Y}{p_Y+p_X}$.

\begin{figure}[ht]
    \centering
    \includegraphics[width=0.4\columnwidth,scale=1]{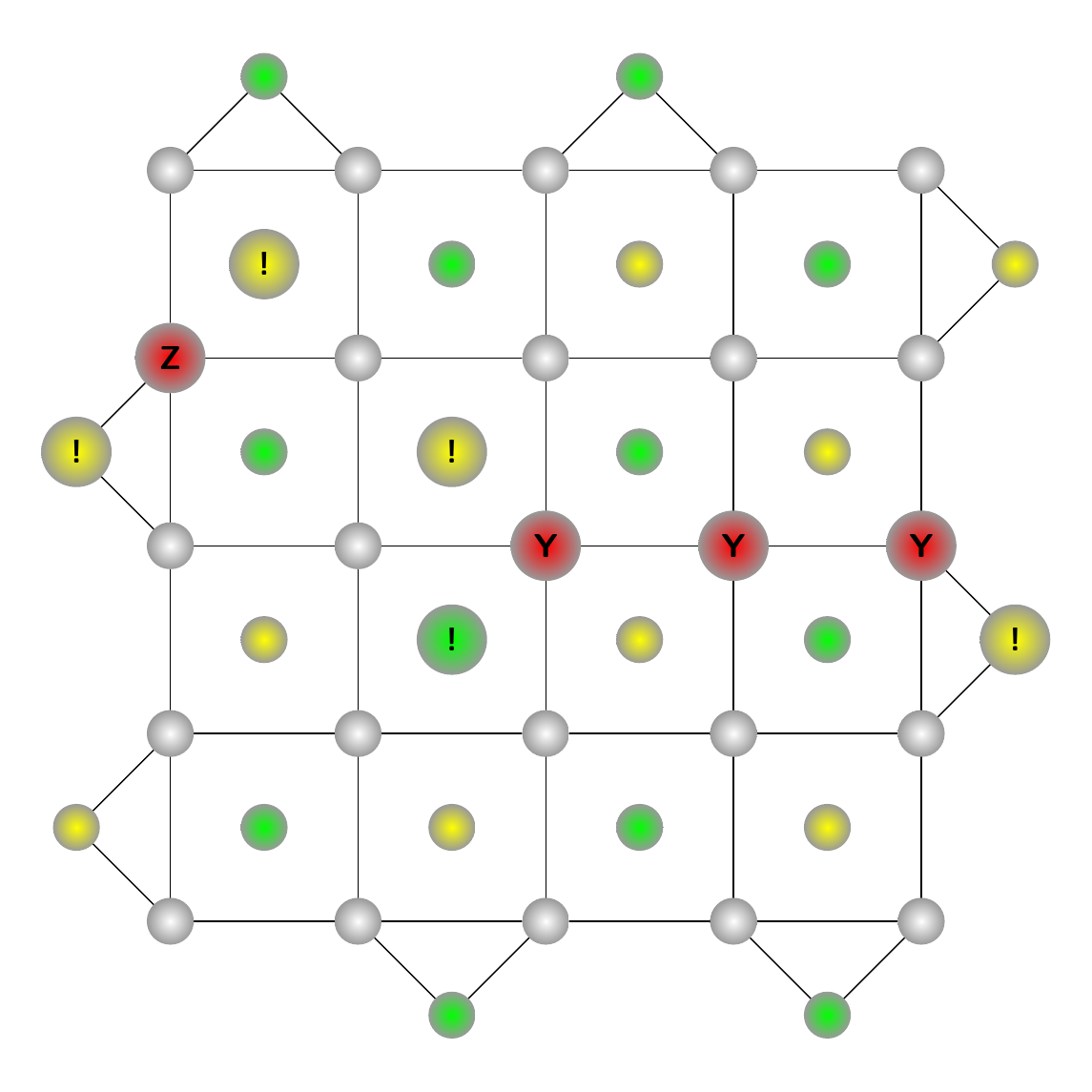}
    \includegraphics[width=0.4\columnwidth,scale=1]{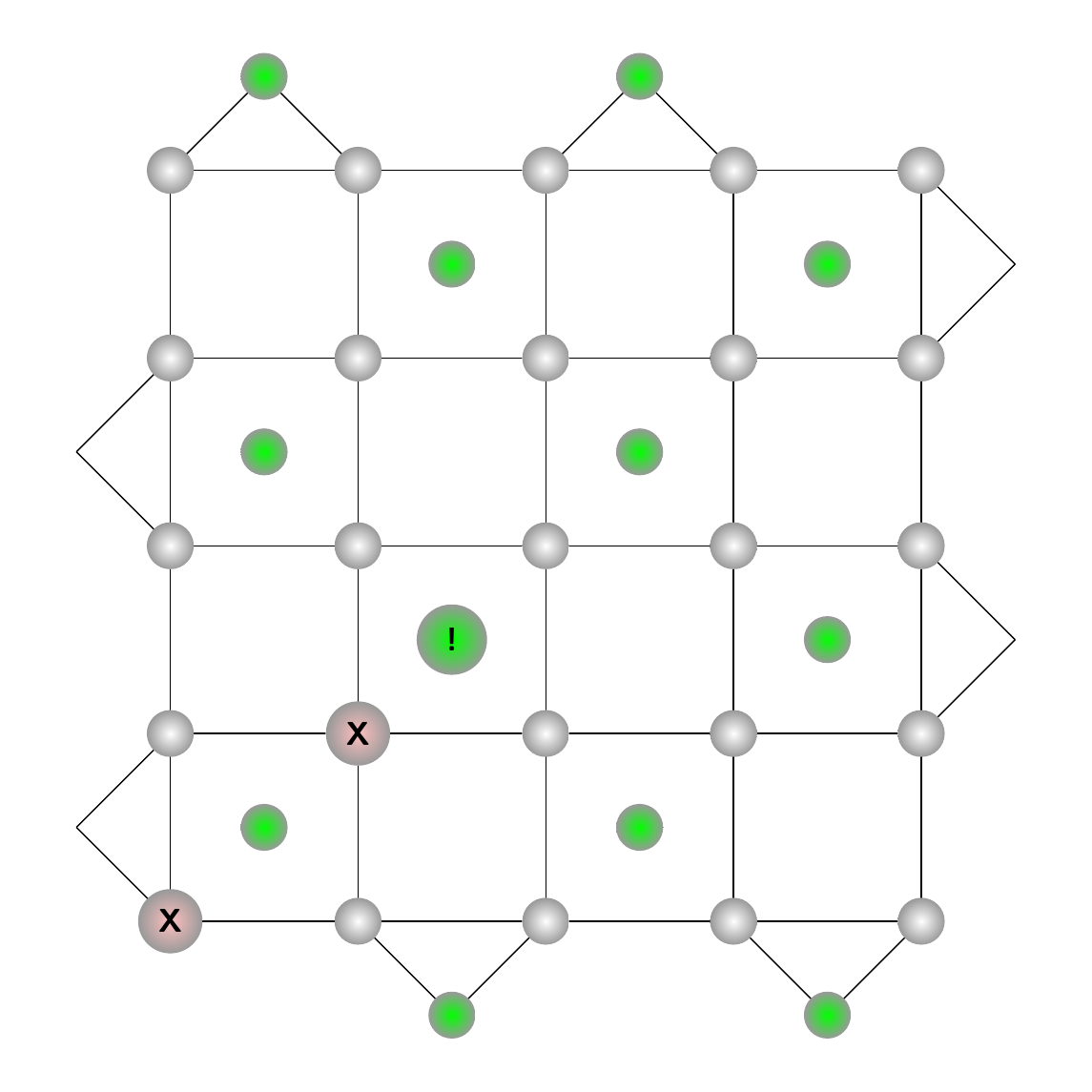}
    \vskip 0.1cm    
    \includegraphics[width=0.4\columnwidth,scale=1]{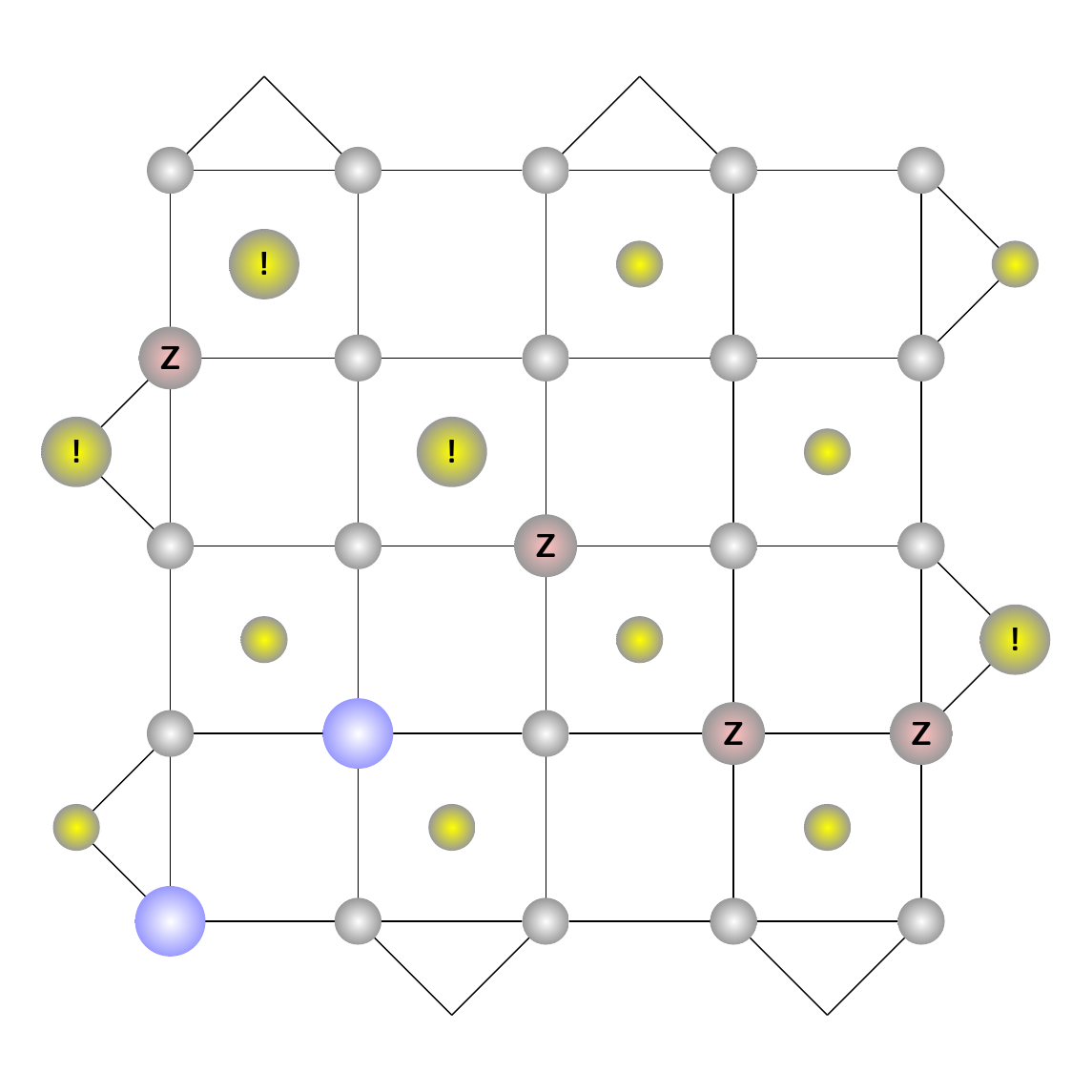}
    \includegraphics[width=0.4\columnwidth,scale=1]{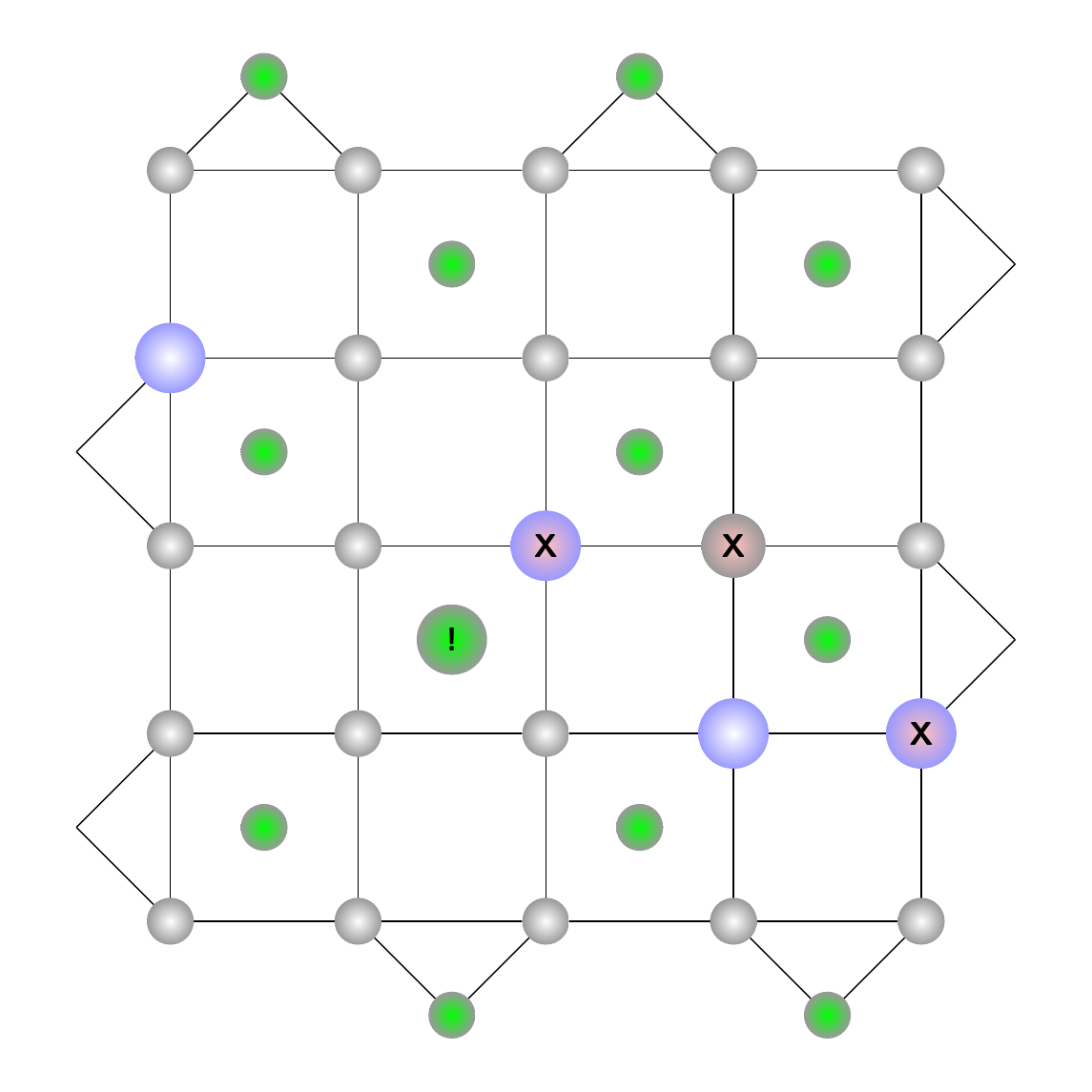}
    \vskip 0.1cm
    \vskip 0.1cm 
    \includegraphics[width=0.4\columnwidth,scale=1]{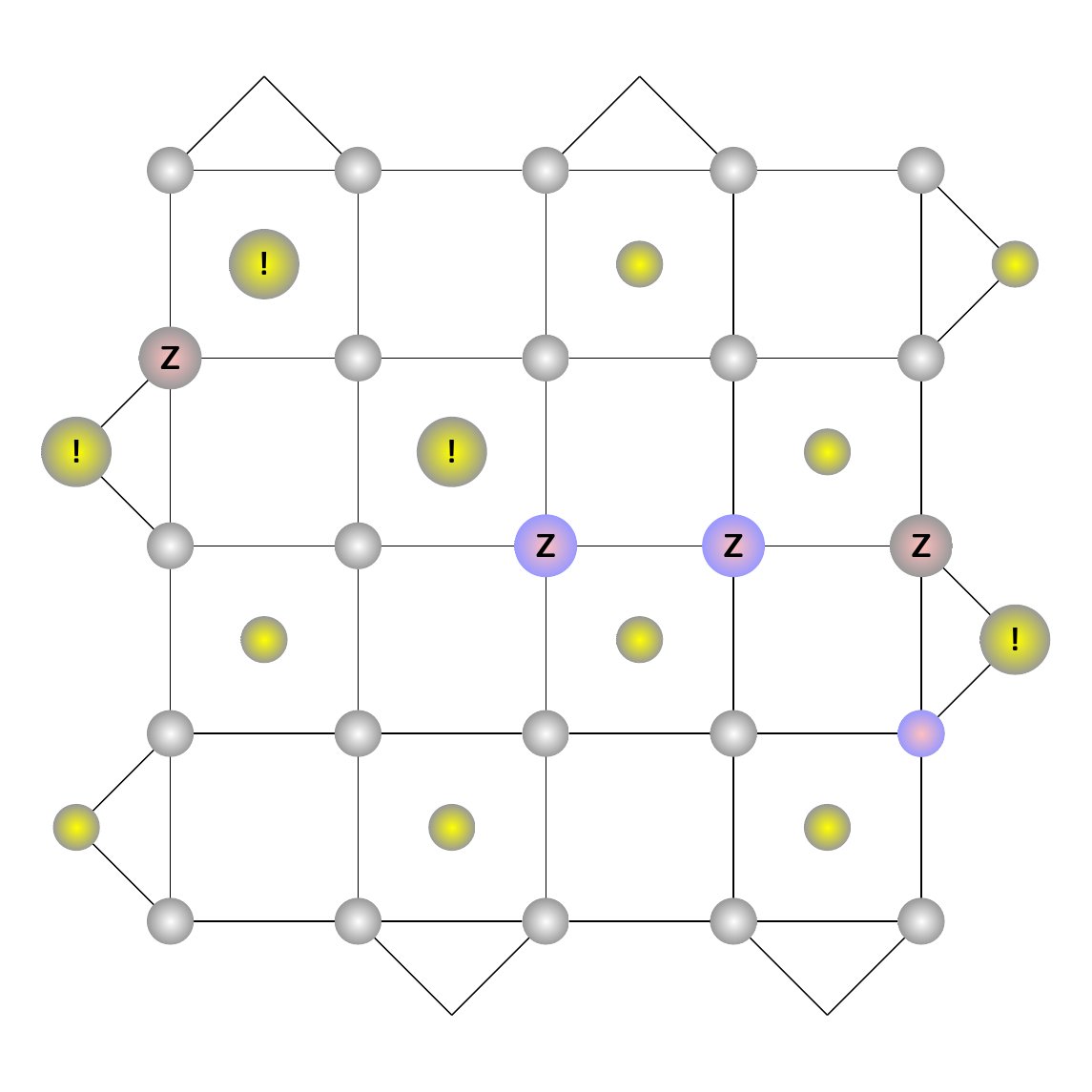}
    \includegraphics[width=0.4\columnwidth,scale=1]{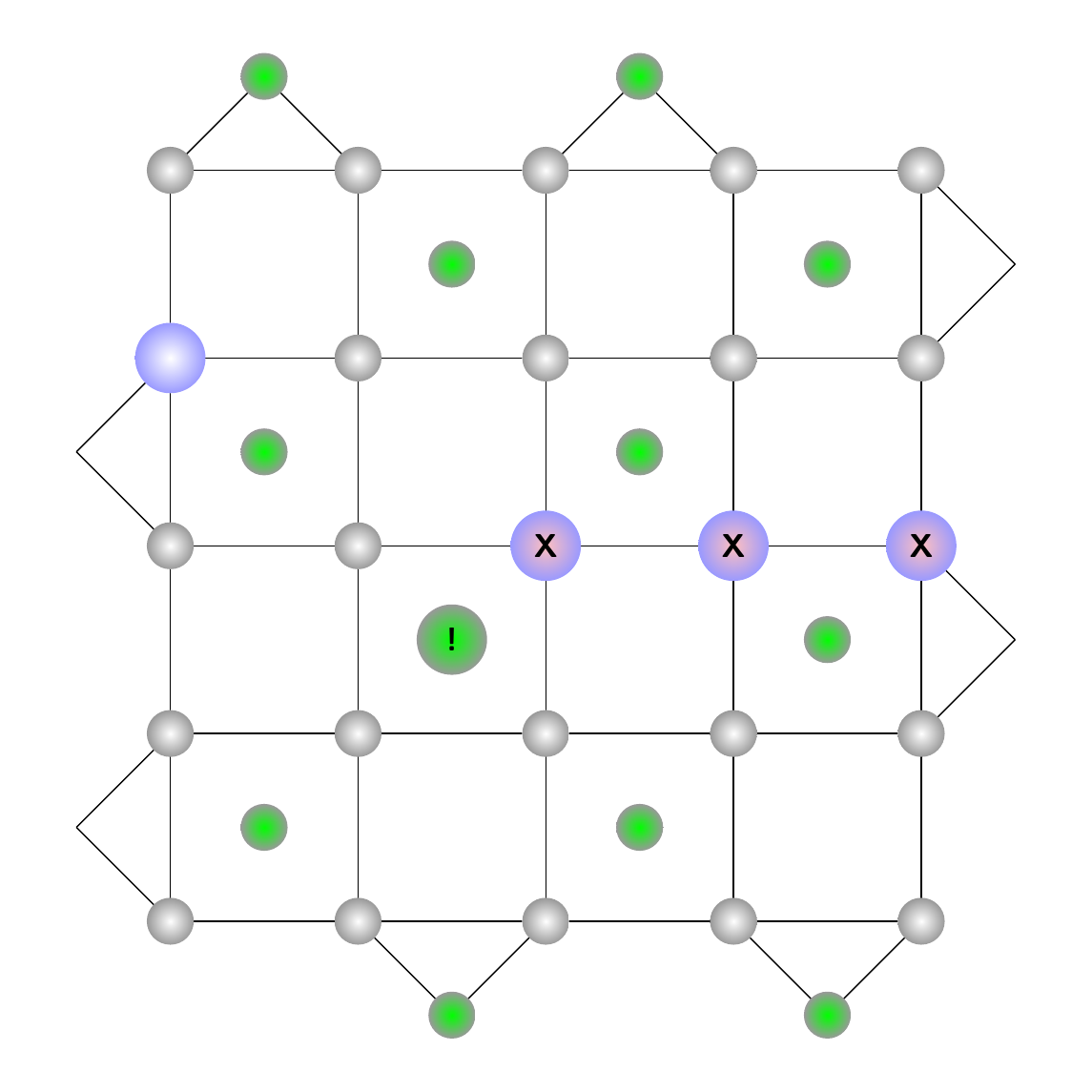}
    \caption{Application of the recursive MWPM strategy on a rotated surface code. White circles represent data qubits, green circles represent X-check qubits and yellow circles represent Z-check qubits. On the \textbf{top left}, the rotated planar code experiences an error (red circles) that causes certain checks to react (exclamation marks). On the \textbf{top right}, the X-check subgraph with the MWPM recovered error (pink circles) is shown. On the \textbf{mid left}, the result of applying the MWPM on the Z-check subgraph can be seen. The blue circles indicate the values chosen for recovery in the previous subgraph, which have changed weight following eq.\ref{logweight}. The remaining figures: \textbf{mid right}, \textbf{bottom left} and \textbf{bottom right}, portray the remaining stages of the recursive MWPM until the error is found. }
    \label{recMWPM}
\end{figure}

Once the matching on the first subgraph is computed, the next step is to re-weigh the data qubits in the second subgraph according to:

\begin{equation}
\label{logweight}
    w_i = -\log(p(Z_i=1 | X_i)).
\end{equation}

\noindent Where, $Z_i=1$ indicates that the $i$th edge in the $Z$-check subgraph recovers an $Z$-error, and $X_i$ indicates the result obtained in the earlier $X$-check subgraph for the $i$-th qubit. That is, the weight of each edge is determined for the probability that an $Z$-error arises knowing the outcome of that same edge in the other subgraph. In the symmetric depolarizing channel, where $p_X = p_Y = p_Z = p/3$:

\begin{equation}
\label{weights}
\begin{aligned}
    p(X=1 | Z=1) &= \frac{p_Y}{p_Z + p_Y} = \frac{1}{2},
    \\
    p(X=1 | Z=0) &= p_X = p/3,
    \\
    p(Z=1 | X=1) &= \frac{p_Y}{p_X + p_Y} = \frac{1}{2},
    \\
    p(Z=1 | X=0) &= p_Z = p/3.
\end{aligned}
\end{equation}

\noindent Once the second subgraph is obtained, the same procedure is repeated by reweighing the first graph following eq.\ref{logweight}. This process is repeated, until the recovered error of a subgraph matches its previous one, i.e, the estimate does not change following two recoveries.

 Fig.\ref{recMWPM} provides a graphical portrayal of the procedure of the recursive MWPM strategy. Once the code experiences an error, first the $X$-check subgraph is computed. Then, a re-weighted $Z$-check subgraph is computed. This process continues until the result of a subgraph occurs twice. It is possible in some instances that this halting criterion is never satisfied. To avoid them, we establish a hard limit of subgraph MWPM computations represented by $N_{max}$. Once this limit is achieved, we reduce the computation to a conventional MWPM problem.

\begin{figure}[h]
    \centering
    \includegraphics[width=0.9\columnwidth,scale=1]{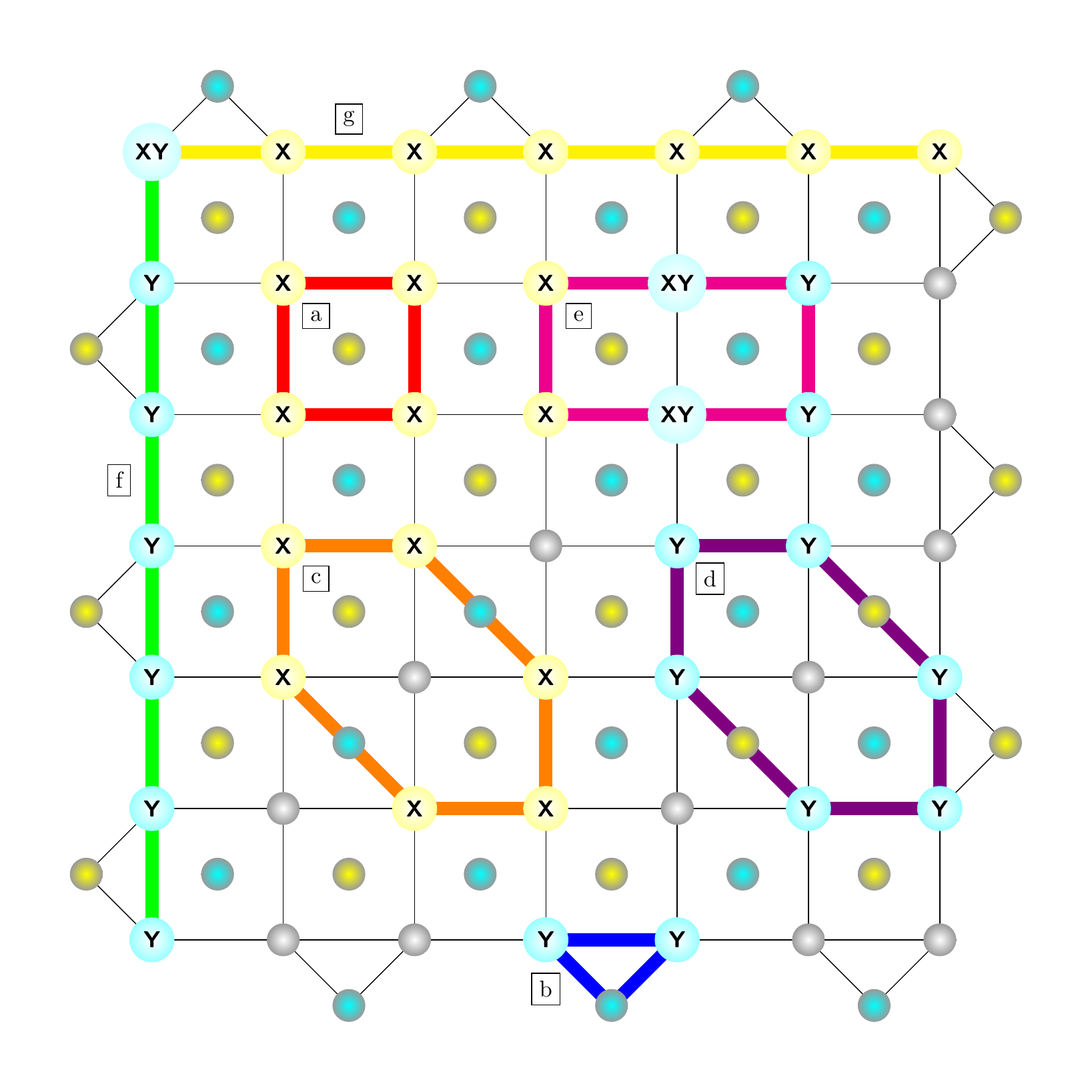}
    \caption{ $7\times7$ rotated planar code with $Y$-measurements instead of $Z$-measurements. }
    \label{xyrotated}
\end{figure}

The recursive MWPM modification increases the overall complexity of the MWPM by $N_{max}+1$ but also significantly increases the robustness of the surface code towards noise which can be detected through both types of checks. In the case of conventional CSS planar codes, recursive MWPM decoding makes the planar code more resistant to $Y$-errors when compared to the conventional MWPM decoder (since both $Z$ and $X$-checks are susceptible to them). We know from the literature that qubits are more prone to suffer from $Z$-errors \cite{supercbias, josurev, TVQC}. In order to preserve the tolerance towards biased noise that we established, we propose changing the checks so as to have a code with $X$-checks and $Y$-checks, where $Z$-operators are detected by both. This technique is commonly known as the XY surface code, and has been used to enhance other codes when facing highly biased noise \cite{tuckbias1, tuckbias2, fragile}. In Fig.\ref{xyrotated}, an example of such a code is provided. The stabilizing properties of the code remain unchanged, but the reaction of the code towards $Y$ and $Z$-operators is reversed. This allows for more shared information between subgraphs and, consequently, improvements in performance when recMWPM is used.

\section{Results}

\begin{figure}[h]
    \centering
    \includegraphics[width=\columnwidth,scale=1]{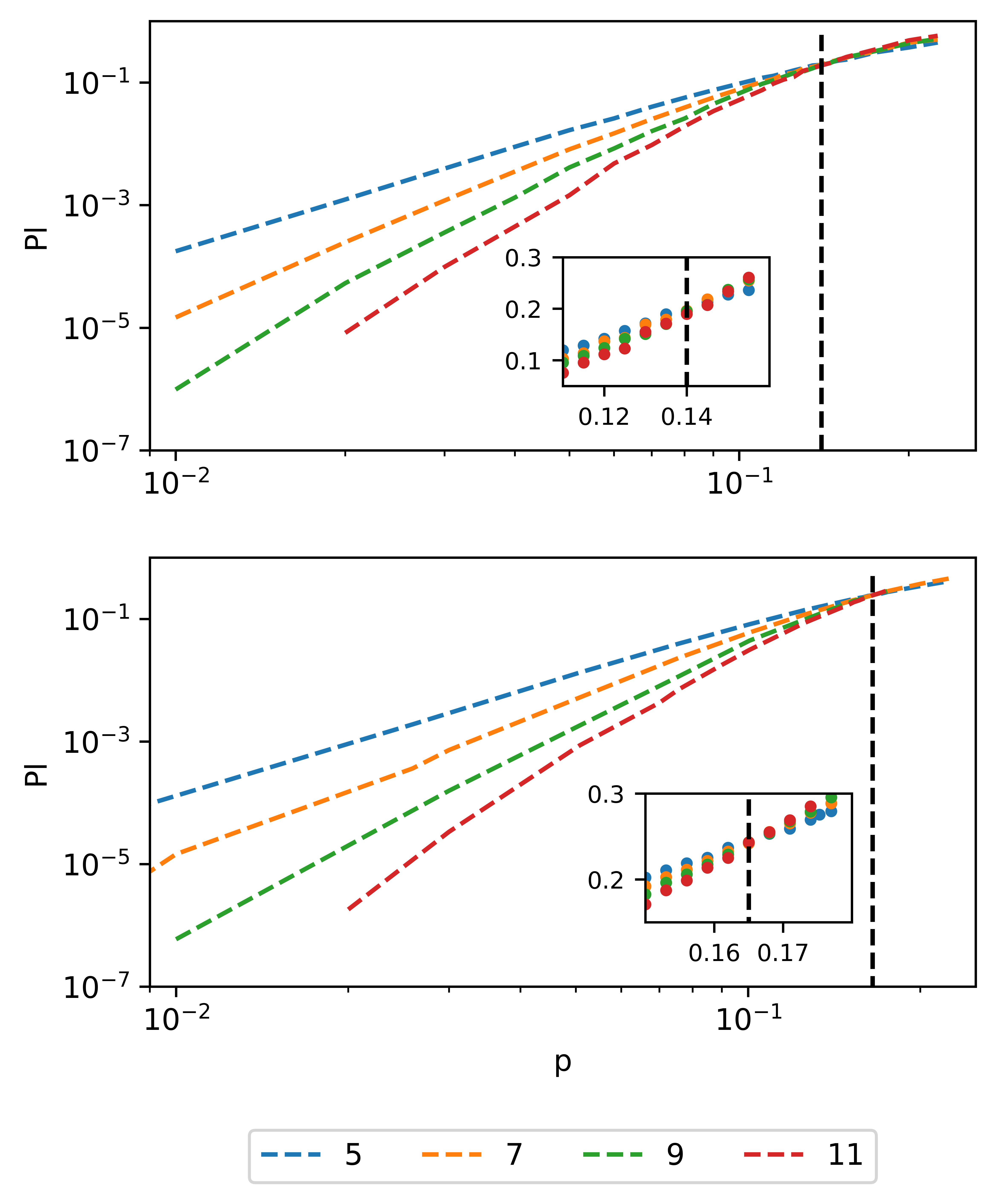}
    \caption{ Logical error rate as a function of the physical qubit error rate of the symmetric depolarizing channel for four different distance rotated planar codes. The top figure shows the results for the conventional MWPM decoding method. The bottom figure portrays the results for the recursive MWPM method. The black dashed lines indicate the location of the threshold.}
    \label{threshold}
\end{figure}

The performance of the recMWPM method used to decode rotated planar codes has been assessed over three different noise models. Over the depolarizing channel ($p_X = p_Y = p_Z = p/3$) the performance of the code has been benchmarked through the probability threshold ($p_{th}$), that is, the physical error probability at which increasing the distance of the code does improve its performance. As can bee seen in Fig.\ref{threshold}, the rotated planar code decoded via recursive MWPM reaches a probability threshold at $p = 16.5\%$. This represents a performance gain of around $18\%$ with respect to the conventional MWPM.

\begin{figure}[h]
    \centering
    \includegraphics[width=0.9\columnwidth,scale=1]{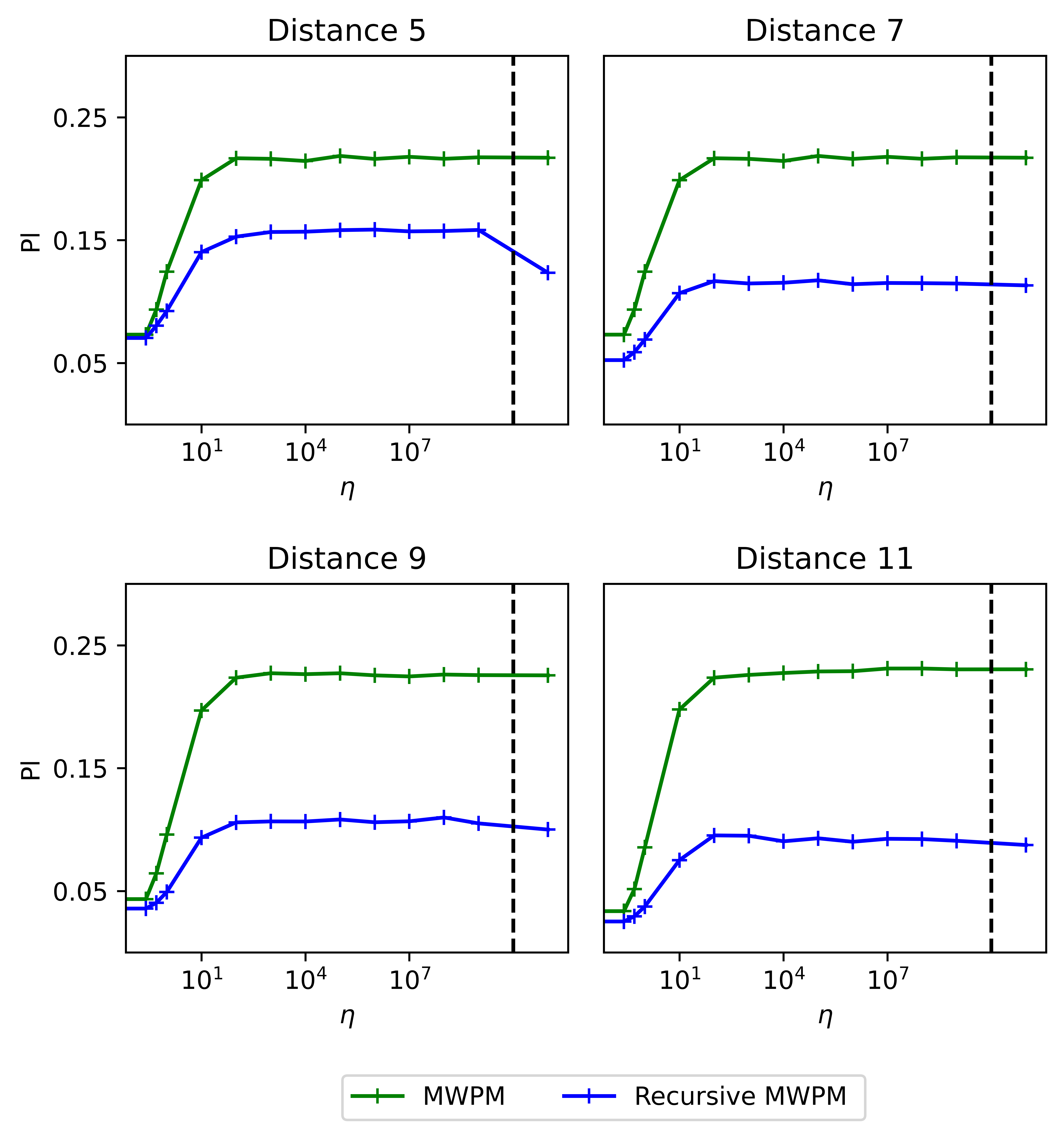}
    \caption{ Logical error rate as a function of $Z$-error bias with a physical error probability of $10\%$ for four rotated planar codes of distances $5, 7 , 9$ and $11$. The green curve represents the probability of error under the conventional MWPM decoding method. The blue curve indicates the performance of the recursive MWPM method. The dashed lines separate the last point, which considers infinite bias (pure $Z$-noise). }
    \label{Plvsbias}
\end{figure}

The strength of the recursive MWPM decoder is showcased further when biased noise is considered. Over the biased channel, qubit errors have independent and identically distributed probabilities but there is a bias towards a specific type of error. In our case we consider bias towards $Z$-noise, which will be determined by $\eta = \frac{p_Z}{p_X+p_Y} = \frac{pz}{2p}$. For a fixed physical error probability of $10\%$ ($p_X + p_Y + p_Z = 0.1$) the performance of the code using MWPM and recMWPM is shown in Fig.\ref{Plvsbias}. As can be seen in said figure, the performance loss experienced when using conventional MWPM is significantly reduced when using recursive MWPM. Additionally, the difference in performance between both decoding methods increases as we use higher distance codes.


Evaluating the performance of the rotated planar code under i.ni.d. is a complicated task. In  the Appendix, it is shown that the relaxation and dephasing times of qubits in state of the art quantum processors vary significantly, up to orders of magnitude \cite{googlesurface, Aspen, Zuchongzhi, Wash}. This experimental phenomenon presents us with two challenges to benchmark the performance of surface codes. The first one is that choosing different sets of qubits with the same average $T_1$ and $T_2$, while possible, will result in different simulated performances. This occurs because since the standard deviations of the different samples may vary, similar qubit samples will outperform samples whose qubits are very different. To resolve this issue, we adopt a strategy where, for each code size, we select the qubits with the longest and shortest relaxation and dephasing times. This approach is taken for maintaining the average relaxation and dephasing times of the qubit set while maximizing the impact of i.ni.d. noise through a substantial standard deviation in the values of $T_1$ and $T_2$. The second challenge revolves around the fact that we can no longer use the probability threshold ($p_{th}$) as our performance reference. As shown in \cite{ton}, the performance curves do not converge in a $p_{th}$ due to the differences in the data qubit samples. Consequently, we will now use the probability pseudo-threshold ($p_{pth}$), that is, the probability at which the logical error probability ($P_L$) equals the physical error probability, i.e. $P_L = p_{pth}$, as our performance reference metric. Said physical error probability is obtained by considering i.i.d. noise for the average relaxation and dephasing times.

\begin{figure*}[th]
    \centering
    \includegraphics[width=1.8\columnwidth,scale=1]{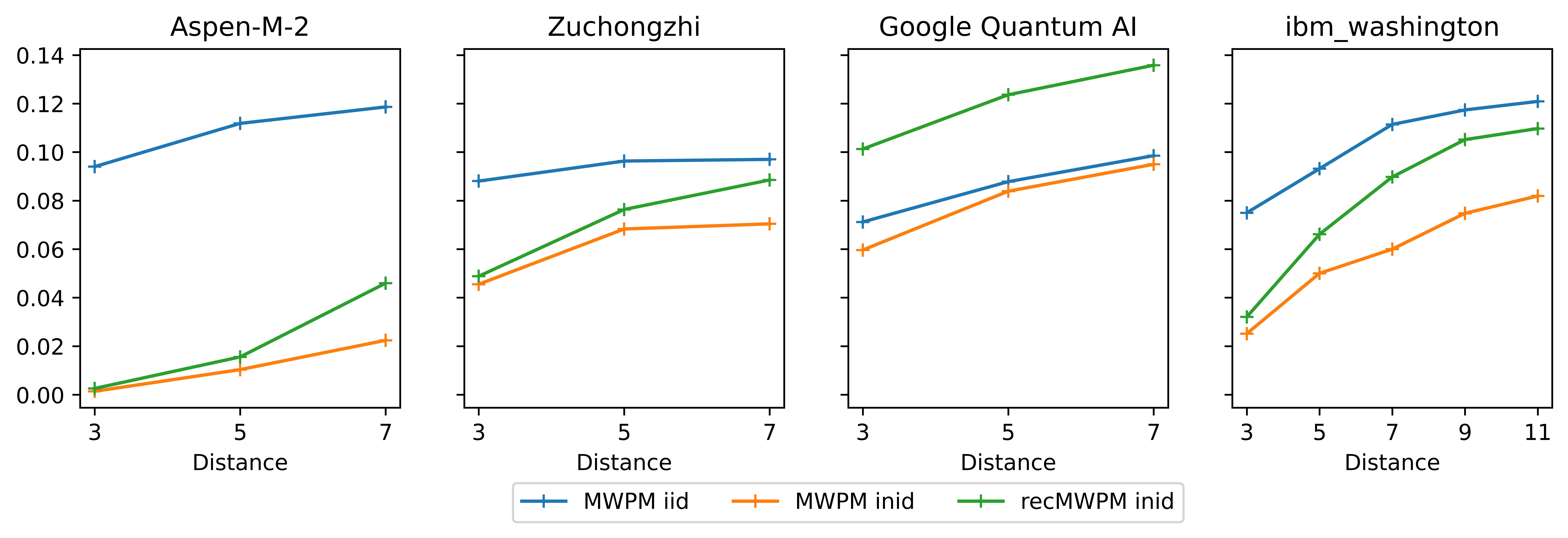}
    \caption{ Probability pseudo-thresholds as functions of the code distances for four different sets of qubits under three different scenarios. The blue curves consider i.i.d. conditions and are decoded through MWPM. The orange and green curves consider i.ni.d. noise decoded with MWPM and recMWPM, respectively.}
    \label{performances}
\end{figure*}

Previous works have shown that considering samples of real computer readout $T_1$ and $T_2$ parameters results in decreases in the performance of the conventional MWPM decoder, that is, MWPM where all edges are equal, when applied to planar codes \cite{ton}. As can be seen in Fig.\ref{performances}, this phenomena is also observed for rotated planar codes. For this work, we have used the $T_1$ and $T_2$ values measured in several state-of-the-art quantum processors in order to simulate performance of rotated planar codes when i.ni.d. noise is considered \cite{googlesurface, Aspen, Zuchongzhi, Wash}. The change between i.i.d. noise to i.ni.d. noise reduces the $p_{pth}$ of the code by $3.6\% $ for a rotated planar code of distance-7 under a noise channel produced by the $T_1$ and $T_2$ values from the Google chip \cite{googlesurface} to $98.5 \%$ in the case of the distance-3 rotated planar code with the data of the the Aspen-M-2 qubits \cite{Aspen}. The performance loss for the smaller code is caused both because of the higher relevance of the worst performing qubits within the code and by the fact that, given the smaller distance, lower weight errors are needed to cause code failure. 

Applying the recMWPM strategy enhances the performance when compared to conventional MWPM methods by $7.2\%$ improvement of the $p_{pth}$ for Zuchongzhi qubits \cite{Zuchongzhi} in a distance-3 code and $105.0\%$ for the distance-7 code with the qubits of Aspen-M-2 \cite{Aspen}. When considering Google processor's qubits \cite{googlesurface}, the recursive MWPM decoder surpasses the MWPM strategy by $42.3\%$ when subjected to i.ni.d. noise. This exceptional performance improvement for the Google quantum processor is explained by the fact that qubits whose parameters satisfy $T_1 > T_2$ and $T_1 < T_2$ are present. This makes it so that the decoding subgraphs that are generated are less dense, which ultimately improves MWPM performance. Ultimately, employing the recMWPM strategy provides an advantage over executing the conventional MWPM decoder under i.ni.d. for all of the simulated distances.

\section{Simulations}

The simulations we have carried out for this paper have been conducted using a modified version of the QECSIM package \cite{qecsim}. For both the symmetric Pauli channel and the biased Pauli channel the logical probability of error has been derived following $10^5$ applications of random errors to the code. For the i.ni.d. results, 100 different qubit distributions (arrangements on the lattice) have been considered for each point and for each specific point 15 probabilities have been considered (the closest one to the $p_{pth}$ taking into account its error was chosen). {For all cases considered $N_{max}$ has been assigned a value of 10. Finally, the weighted average has been employed to obtain the results.

\section{Conclusion} \label{sec:conclusion}

This work introduces an improved decoding strategy for surface codes known as the recursive minimum weight perfect matching decoding method. We have shown how considering the decoding outcomes of one subgraph to condition and solve the next one can yield improvements in the performance of the decoder. Our simulation results show an $18\%$ increase in the threshold of the rotated planar code under depolarizing noise when using this recursive decoder. We have also shown how the recMWPM strategy is more robust towards biased noise than the generic MWPM decoder. Finally, we also studied the behaviour of our decoding approach when subjected to i.ni.d. noise. As expected, we observed a decrease in the performance similar to the one experienced by the conventional MWPM decoder over such channels. Nonetheless, our results prove that the recMWPM method also surpasses the MWPM decoder under i.ni.d. noise. Additionally, although in this work the decoding method has only been applied to the rotated planar code, it could be successfully extrapolated to the toric and planar codes by adapting the the subgraph structures.

The primary takeaway from this work is that small changes to the MWPM decoding method can result in strong performance improvements when certain information about qubit noise is known. The recursive minimum weight perfect matching method proposed herein is an example of this, as it achieves significant improvements in performance at the expense of small increases in decoding complexity. 

Reweighting the edges of subgraphs for MWPM computation is not a new task. In \cite{fragile}, edges are adapted based on the resulting probabilities of a BP process. In \cite{xzcorrelation}, one of the subgraphs is decoded using conventional MWPM, and the second subgraph is reweighted based on the earlier matching. In \cite{fowlercorr}, single matches within the bulk of the code are used to reweight the edges of the other subgraph. In contrast, recMWPM continuously updates both subgraphs until they converge to an error. In \cite{pipeline}, errors within the stabilizer circuit of the checks are considered to reweight the edges. For recMWPM, measurement gates are assumed to be perfect, and the correlations between the $X$ and $Z$-subgraphs are the only source for reweighting. In \cite{oldrec}, a recursive MWPM method is performed under depolarizing noise, but matched edges in one subgraph have weight-0 while the remaining have weight-1 in the other subgraph. RecMWPM adapts the weight to the channel conditions and, more specifically, to the individual qubit's conditions, as shown in equation \ref{logweight}. This difference allows it to better undergo biased channels and i.ni.d. noise.

Furthermore, these conclusions give rise to some crucial unresolved questions. One critical factor in the recMWPM algorithm is the hard limit parameter, $N_{max}$. A higher value of $N_{max}$ may lead to improved code performance, especially in the case of independent and non-identical scenarios where qubits' $T_1$ and $T_2$ values differ significantly and different paths have very different weights. However, a larger $N_{max}$ also results in a higher decoding complexity, while a lower $N_{max}$ may not fully exploit the correlation of the subgraphs to an optimal extent. This becomes especially important since measurement errors may occur implying that several stabilizing circuits might be necessitated before the overall decoding process. This increases the complexity of the decoding process even more, and this must to be done in real time. Future research may focus on how to apply the recMWPM algorithm to a decoding scheme with measurement error and how the qubits' constraints may restrict the maximum value of $N_{max}$.
}

\section*{Data availability}
The data that supports the findings of this study is available from the corresponding authors upon reasonable request.

\section*{Code availability}
The code that supports the findings of this study is available from the corresponding authors upon reasonable request.

\section*{Author Contributions}
A.dM.iO., P.F. and J.E.M. conceived the research and proposed the recMWPM method. A.dM.iO. performed the numerical simulations. A.dM.iO., J.E.M., and P.F. analyzed the results and drew the conclusions. The manuscript was written by A.dM.iO., P.F., and J.E.M., and revised by P.M.C. The project was supervised by P.M.C.

\section*{Competing Interests}
The authors declare no competing interests.

\section*{Acknowledgements}
This work was supported by the Spanish Ministry of Economy and Competitiveness through the ADELE project (Grant No. PID2019-104958RB-C44), by the Spanish Ministry of Science and Innovation through the project Few-qubit quantum hardware, algorithms and codes, on photonic and solid-state systems (PLEC2021-008251), by the Ministry of Economic Affairs and Digital Transformation of the Spanish Government through the QUANTUM ENIA project call - QUANTUM SPAIN project, and by the European Union through the Recovery, Transformation and Resilience Plan - NextGenerationEU within the framework of the Digital Spain 2025 Agenda, and by the Diputación Foral de Gipuzkoa through the Post-Quantum Cryptographic Strategies for Critical Infrastructure project.

\appendix\section{ The independent non-identical noise model for superconducting qubits}

At the time of writing, most of the accessible experimental quantum computing prototypes have been built using superconducting qubits. Superconducting qubits generally suffer from noise that stems from the combination of two processes: energy relaxation and pure dephasing. Energy relaxation describes the process of energy loss in a qubit arising from the emission of a photon, while pure dephasing refers to a change of phase within the qubit. The Lindblad master equation provides a mathematical description of both these phenomena through the amplitude damping channel $\mathcal{N}_{\mathrm{AD}}$ and the dephasing channel $\mathcal{N}_{\mathrm{PD}}$, respectively. To study both effects simultaneously, we can make use of the amplitude and phase damping channel $\mathcal{N}_{\mathrm{APD}}$, which combines both the effects of dephasing and relaxation. Unfortunately, the amplitude and phase damping channel is costly to compute, since the overall dimension of the Hilbert space of $n$-qubits scales as $2^n$. To efficiently simulate this type of quantum noise we apply the information-theoretic concept known as twirling. Twirling is a widely used method in classical and quantum information which consists in studying the average effect of noise by mapping it to a more symmetric version of itself \cite{josurev, TVQC}. The application of a Pauli twirl approximation to the amplitude and phase damping channel results in the Pauli channel:

\small
\begin{equation}
    \mathcal{N}_{\mathrm{APDPTA}}(\rho) = (1-p_X-p_Y-p_Z)\rho + p_XX\rho X+ p_YY\rho Y+ p_ZZ\rho Z,
\end{equation}

\noindent where $I$, $X$, $Y$ and $Z$ are the identity, bit-flip, bit and phase-flip and phase-flip Pauli matrices, respectively. Moreover, $p_X$, $p_Y$ and $p_Z$ are the probabilities of each of the non-trivial processes and correspond to:

\begin{equation}\label{eq:PTAprobs}
\begin{split}
& p_{X} = p_{Y} = \frac{1}{4}(1 - \mathrm{e}^{-\frac{t}{T_1}})\text{ and} \\
& p_{Z} = \frac{1}{4}(1 + \mathrm{e}^{-\frac{t}{T_1}} - 2\mathrm{e}^{-\frac{t}{T_2}}).
\end{split}
\end{equation}

\noindent Based on this, it is clear that the noise suffered by the qubits that comprise our surface codes is governed by the relaxation time ($T_1$) and the dephasing time ($T_2$). Qubits limited by their $T_1$ will be more prone to $X$ and $Y$-errors, while the ones limited by $T_2$ will be more susceptible towards $Z$-errors. The most studied quantum channel in the literature is the depolarizing channel, that is, the channel where $T_1 = T_2$ and so $p_X = p_Y = p_Z$. However, it is likely that superconducting qubits are more biased towards $Z$-noise, hence, great effort has been expended recently in building codes tailored to such noise profiles.

Additionally, much of the current research on quantum codes considers all qubits within the code to have the same $T_1$ and $T_2$ parameters, which means that they all have independent and identically distributed error probabilities (i.i.d.). This means, that for an $n$-qubit system, the Pauli twirl approximation amplitude and phase damping channel is described by:

\begin{equation}\label{eq:iidchan}
\begin{split}
\mathcal{N}^{(n)}_\mathrm{APDPTA}(\rho) &= \mathcal{N}_\mathrm{APDPTA}^{\otimes n}(\rho,\mu_{T_1},\mu_{T_2})\\  &= \sum_{\mathrm{A}\in\{\mathrm{I,X,Y,Z}\}^{\otimes n}} p_\mathrm{A}(\mu_{T_1},\mu_{T_2}) \mathrm{A}\rho\mathrm{A},
\end{split}
\end{equation}

\noindent where $\mathrm{A} = \mathrm{A}_{1}\otimes\cdots\otimes\mathrm{A}_{n-1}\otimes\mathrm{A}_n$ with $\mathrm{A}_i\in\{{I,X,Y,Z}\}$ denotes each of the possible $n$-fold Pauli error operators, with probability distribution $p_\mathrm{A}(\mu_{T_1},\mu_{T_2})$
\begin{equation}\label{eq:iidprob}
p_{\mathrm{A}}(\mu_{T_1},\mu_{T_2}) = \prod_{i=1}^{n}p_{\mathrm{A}_i}(\mu_{T_1},\mu_{T_2}),
\end{equation}

\noindent with $p_{\mathrm{A}_j}(\mu_{T_1},\mu_{T_2})$ described by equation \eqref{eq:PTAprobs}, and where $\mu_{T_1}$ and $\mu_{T_2}$ represent the mean values of the relaxation and dephasing times averaged across $n$ qubits.

In reality, measurements on state of the art quantum processors have shown that the $T_1$ and $T_2$ parameters vary significantly from qubit to qubit, sometimes even by an order of magnitude. This inter-qubit change is too significant to be ignored and is not conveyed by taking the average of the time parameters, thus, a new model must be derived by considering each data qubit separately:

\begin{equation}\label{eq:inidchan}
\begin{split}
\mathcal{N}^{(n)}_\mathrm{APDPTA}(\rho) &= \bigotimes_{i=1}^n\mathcal{N}_\mathrm{APDPTA}(\rho,\mu_{T_1^i},\mu_{T_2^i})\\  &= \sum_{\mathrm{A}\in\{\mathrm{I,X,Y,Z}\}^{\otimes n}} p_\mathrm{A}(\{T_1^i\}_{i=1}^n,\{T_2^i\}_{i=1}^n) \mathrm{A}\rho\mathrm{A},
\end{split}
\end{equation}
where $\mathrm{A} = \mathrm{A}_{1}\otimes\cdots\otimes\mathrm{A}_{n-1}\otimes\mathrm{A}_n$ with $\mathrm{A}_i\in\{{I,X,Y,Z}\}$ denotes each of the possible $n$-fold Pauli error operators with probability distribution $p_\mathrm{A}(\{T_1^i\}_{i=1}^n,\{T_2^i\}_{i=1}^n)$
\begin{equation}\label{eq:inidprob}
p_\mathrm{A}(\{T_1^i\}_{i=1}^n,\{T_2^i\}_{i=1}^n) = \prod_{i=1}^{n}p_{\mathrm{A}_j}(T_1^i,T_2^i),
\end{equation}

\begin{figure}[h!]
    \centering
    \includegraphics[width=0.95\columnwidth,scale=1]{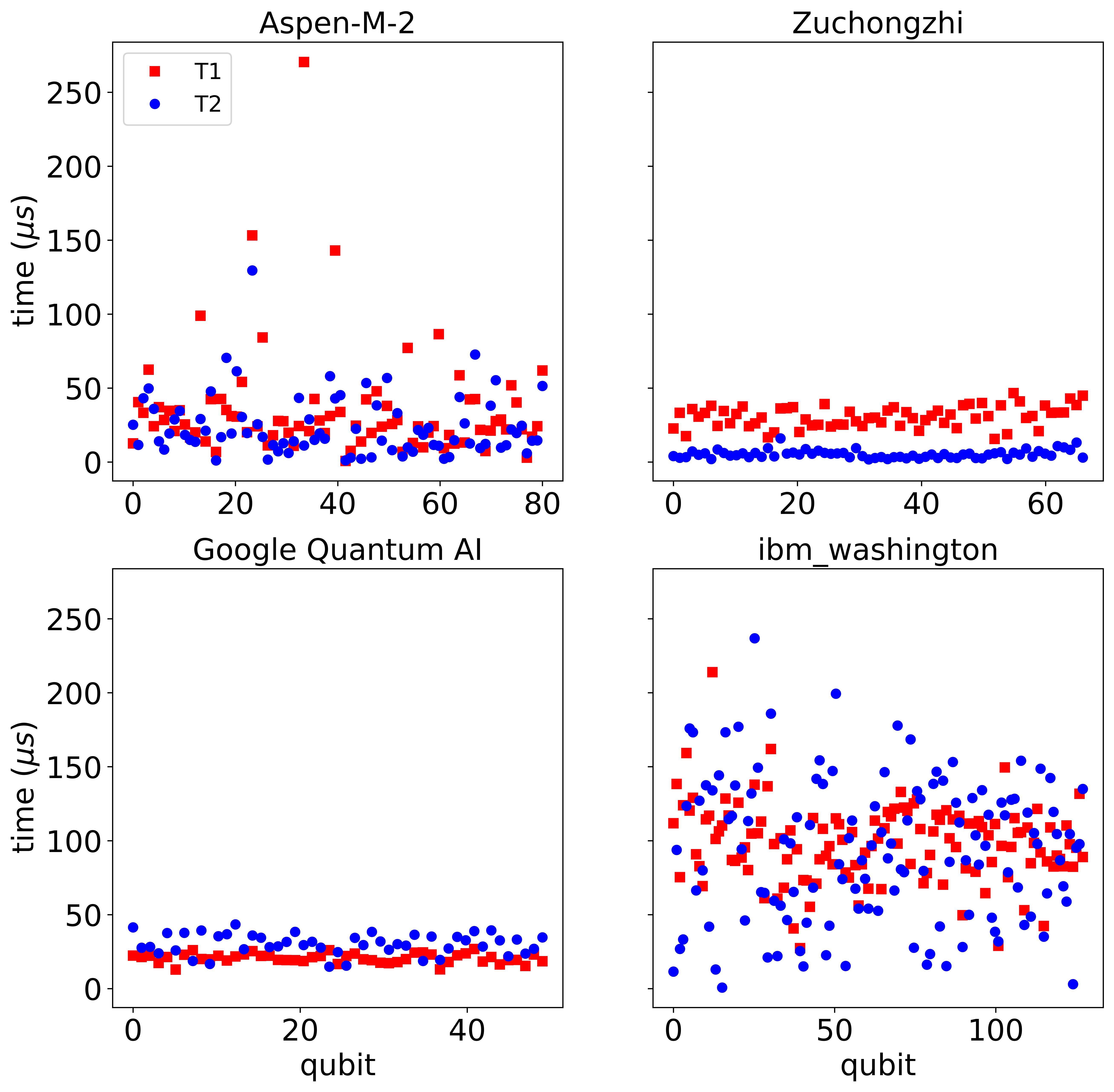}
    \caption{ Representation of the values of $T_1$ and $T_2$ from the qubits in the quantum processors used for simulations.}
    \label{t1st2s}
\end{figure}

This independent consideration of the qubit parameters is named independent non-identically distributed error model (i.ni.d.) \cite{ton, fastfading}. In FIG.\ref{t1st2s}, a representation of the relaxation and dephasing times of the qubits from the chosen quantum processors is portrayed. The experimental measurements reveal how the particular relaxation and dephasing times of each individual qubit within the quantum system can vary drastically. For example, the qubits that make up the ibm\_washington quantum processor exhibit a minimum relaxation time of $16.54$ $\mu s $ and a maximum relaxation time of $123.11$ $\mu s$, i.e, there are qubits whose relaxation time differs by an order of magnitude. This phenomenon is further exacerbated for the ibm\_washington qubit dephasing times. The minimum dephasing time value is $8.58$ $\mu s$ and the maximum value is $228.56$ $\mu s$. This behaviour can be relatively observed over all of the superconducting machines considered in this paper. The main takeaway here is that, within the real quantum systems, the decoherence parameters of each constituent qubit will vary significantly. Because this type of behaviour must be considered when building accurate decoherence models, the i.ni.d. noise model is a relevant contribution to the field of QEC, as it can accurately re-enact the real quantum noise processes that experimental multi-qubit systems can suffer.


\end{document}